\documentclass[aps,twocolumn,amsmath,amssymb]{revtex4}
\usepackage{graphicx}
\usepackage{dcolumn}
\usepackage{bm}

\begin{document}

\title{Magnetic field dependent transmission phase of a double dot system in a quantum ring}

\author{M. Sigrist,$^1$ A. Fuhrer,$^1$ T. Ihn,$^1$ K. Ensslin,$^1$ S. E. Ulloa,$^{1,2}$ W. Wegscheider,$^3$ M. Bichler$^4$}
\affiliation{
$^1$Solid State Physics, ETH Z\"urich, 8093 Z\"urich, Switzerland\\
$^2$Department of Physics and Astronomy, Ohio University, Athens, Ohio 45701-2979\\
$^3$Institut f\"ur experimentelle und angewandte Physik, Universit\"at Regensburg, Germany\\
$^4$Walter Schottky Institut, Technische Universit\"at M\"unchen, Germany
}

\date{\today}

\begin{abstract}
The Aharonov-Bohm effect is measured in a four-terminal open ring geometry based on a Ga[Al]As heterostructure. Two quantum dots are embedded in the structure, one in each of the two interfering paths. The number of electrons in the two dots can be controlled independently. The transmission phase is measured as electrons are added to or taken away from the individual quantum dots. Although the measured
phase shifts are in qualitative agreement with theoretical
predictions, the phase evolution exhibits unexpected dependence on the magnetic field. For example, phase lapses are found only in certain ranges of magnetic field.
\end{abstract}

\maketitle

In mesoscopic systems the conductance and the quantum mechanical transmission of charge carriers between contacts are intimately related.
The corresponding transmission amplitude is
a complex number with magnitude and phase.
The phase difference of interfering paths in coherent quantum rings
can be detected in the conductance by measuring Aharonov-Bohm (AB) oscillations. Keeping the transmission phase of one of the interfering paths constant (reference path), the phase change of the other can be measured. This technique opens the attractive possibility to investigate the transmission phase which contains information about the system complementary to the transmission probability.
Multiterminal devices are required for such experiments in order to circumvent the generalized Onsager symmetry relations~\cite{Buttiker86} restricting the transmission phase to values of 0 or $\pi$ in two-terminal devices.

Ring and ring-like interference geometries have been exploited for many types of experiments~\cite{Yacoby95,Buks96,Schuster97,Ji00,Ji02,Buks98,Sprinzak00,Kobayashi02,Holleitner01,Holleitner02,Fuhrer01}. The electronic phase in particular was studied in
a series of pioneering publications~\cite{Yacoby95,Buks96,Schuster97,Ji00,Ji02}. The
transmission phase through a two-terminal device with a single quantum dot has been investigated \cite{Yacoby95} and it was concluded that electron transport through a Coulomb
blockaded quantum dot is at least partially phase coherent.
The phase of the reflection coefficient of a quantum dot in the integer quantum Hall regime was measured in Ref.~\onlinecite{Buks96}. Later, the transmission phase of a quantum dot was studied in a multi-terminal configuration \cite{Schuster97}. It was found that the transmission phase changes by about $\pi$ across a typical Coulomb blockade resonance, as expected. Between
resonances the experiment exhibits so-called phase lapses rather than the  almost constant phase expected in the simplest model.
Similar experiments were carried out on a Kondo-correlated system in Refs.~\onlinecite{Ji00} and~\onlinecite{Ji02}. 

A large number of theoretical papers (for a review see~\cite{Hackenbroich01}) has addressed the issue of the phase lapses and suggested
possible explanations. It was argued that the transmission phase has to be distinguished from the Friedel phase which fulfills a general sum rule and is a monotonous function of energy, while the transmission phase can show phase lapses of $\pi$ when the transmission goes through zero \cite{Lee99,Buttiker99,Buttiker00}.
It was also pointed out that the measured phase may depend on the details of the entire interferometer,
on interactions~\cite{Koenig02}, and on the dot-lead coupling~\cite{Silva02}.
The
phase determined experimentally is the transmission phase through the dot, only if the system is unitary~\cite{Entin02}, i.e. it couples negligibly to its
environment. A tunneling-Hamiltonian approach predicted the phase to be influenced by the number and width of leads connecting to the ring~\cite{Aharony02,Gefen02}. In a scattering matrix approach, however, the limit of many
occupied modes in the channels reflects the
energy dependence of the resonance phase shift~\cite{Weidenmuller02}. The equivalence
of the tunneling-Hamiltonian formulation and the scattering
matrix approach are discussed in~\cite{Kubala02}.
Theoretical proposals have considered the situation of two
quantum dots in an AB interferometer and the possible coupling of
the dots via phase coherent transport through the leads~\cite{Konig02,Kubala02b}. 
In the context of spintronics two coupled
quantum dots in an AB interferometer may be used for detecting entanglement of spin states~\cite{Loss00}.

Here we investigate the phase evolution of a system of
two quantum dots with negligible mutual electrostatic interaction embedded in two arms of a
four-terminal AB ring. Phase measurements in this unique system are desirable, because both arms of the ring including the two dots can be tuned individually, but the phase coherence of the entire system provides an inherent coupling mechanism~\cite{Konig02} and makes transport through the dots interdependent. The arrangement enables us to control and analyze the effect of the interferometer reference arm on
the measured phase experimentally.
The phase evolution of the system is studied when a single electron is added to either of the two quantum dots, to both quantum dots simultaneously, or when an electron is transferred from one dot to the other. The observed phase shifts agree qualitatively with theoretical expectations at elevated magnetic fields. Unexpectedly, the phase evolution is found to depend on the magnetic field and occasional phase lapses occur in certain field ranges. These observations highlight the need to consider nonlocal coherent effects in the entire system.

\begin{figure}[b]
\centering
\includegraphics[width=8cm]{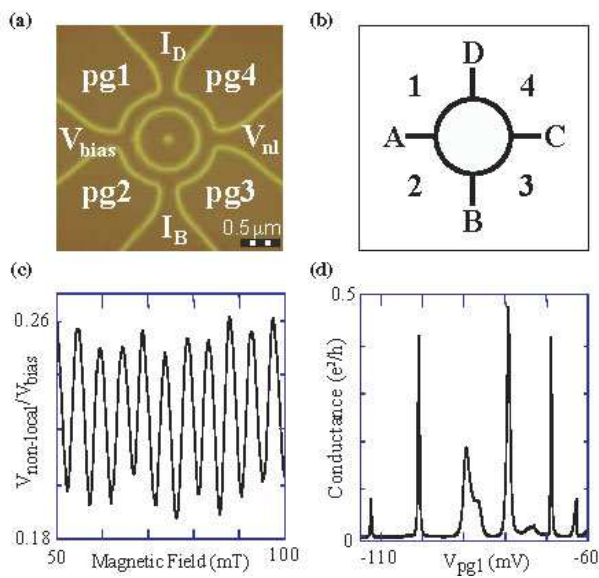}
\caption{\label{Fig1} (a) AFM micrograph
of the ring structure. Oxide lines (bright lines) fabricated by AFM
lithography lead to insulating barriers in the two-dimensional electron gas. Areas marked pg1-pg4 are used as lateral gates to tune the conductance of the four arms of the ring and eventually act as
plunger gates for the dots. (b) Schematic arrangement of the ring with its four terminals (c) Non-local voltage detected in the
open regime of the ring. (d) Conductance through a quantum dot
induced in segment~1 as a function of plunger gate voltage.
Here, segment~3 was completely pinched off.}
\end{figure}
The sample is a Ga[Al]As heterostructure containing a
two-dimensional electron gas 37 nm below the surface. The lateral
pattern was fabricated with the biased tip of an atomic force
microscope which locally oxidizes the GaAs surface. Details of this fabrication technique are described in
Ref.~\onlinecite{Held99}. Figure~\ref{Fig1}(a) shows a micrograph of
the oxidized pattern.  Lateral gate electrodes marked pg1
through pg4 are used to tune the conductance in each of the four
quadrants of the ring. All measurements were carried out at 100\,mK in a dilution refrigerator.

The ring was characterized in the open
regime where each quadrant supports 2-4 lateral modes. A
pronounced AB effect is observed in a local measurement setup
in which the current and voltage contacts are the same two terminals
of the ring \cite{Sigrist03}. The AB signal is maximized with a
non-local setup: a bias voltage $V_\mathrm{bias}$ was applied to terminal A [see Fig.~\ref{Fig1}(b)]. The lower and
upper contacts (B and D) were grounded via current-voltage
converters measuring the currents $I_{\mbox{\tiny B}}$ and $I_{\mbox{\tiny D}}$. The non-local voltage $V_\mathrm{nl}$ was measured at terminal C.
Figure~\ref{Fig1}(c) shows 
the AB oscillations in $V_\mathrm{nl}$. Their period  $\Delta B=\Phi_0 /A=4.8$~mT is consistent with the area $A=0.85$~$\mu$m$^2$ of the ring.

If any individual segment is tuned close to pinch off with a negative voltage applied to the corresponding plunger gate, a
quantum dot is induced exhibiting Coulomb blockade.
As an example, Fig.~\ref{Fig1}(d) shows the conductance of
segment 1 [A to D in
Fig.~\ref{Fig1}(b)] as a function of $V_\mathrm{pg1}$ with
segment~3 completely pinched off and segments 2 and 4 open. Clear Coulomb blockade oscillations are observed.
The location of the quantum dot can be
estimated from the lever arms of gates pg1 to pg4. The dots form within the segments and not at the openings to the contacts. From Coulomb blockade diamonds measured as a function
of forward bias we find a typical Coulomb charging energy of 
1~meV, which corresponds
in a disc capacitor model roughly
to the area of a single segment. More details about the characterisation of the ring can be found in Ref.~\onlinecite{Sigrist03}.

\begin{figure}[b]
\centering
\includegraphics[width=8cm]{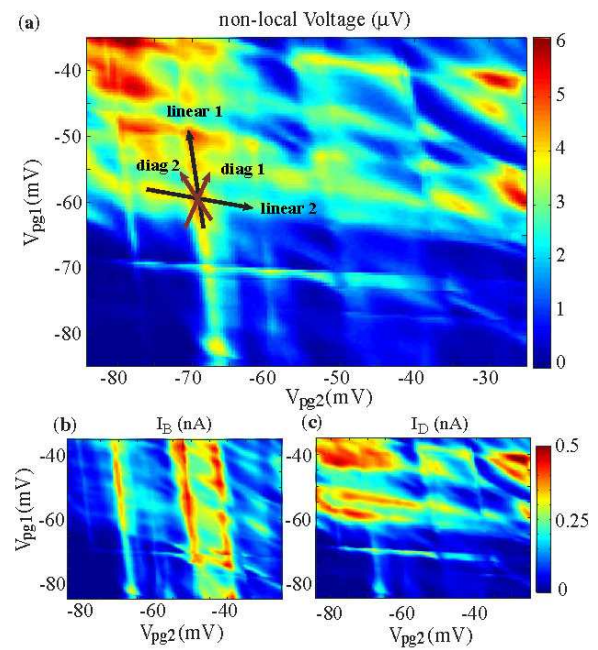}
\caption{\label{Fig2} Color plots of (a) the non-local voltage $V_\mathrm{nl}$, (b) the current $I_{\mbox{\tiny B}}$ and (c) $I_{\mbox{\tiny D}}$
as a function of the two plunger gate
voltages $V_\mathrm{pg1}$ and $V_\mathrm{pg2}$ which tune the quantum dots in segments 1 and 2 of the
ring.}
\end{figure}
We proceed to study Coulomb blockade in the measurement configuration
of Fig.~\ref{Fig1}(a). For the following experiments we
have carefully chosen a series of conductance peaks of one well defined quantum dot in segment~1 (dot~1) and another one in segment~2 (dot~2). Segments~3 and 4 are open. The current
signals $I_{\mbox{\tiny B}}$ and $I_{\mbox{\tiny D}}$ are shown in color as a function of the gate voltages $V_\mathrm{pg1}$ and $V_\mathrm{pg2}$ in Figs.~\ref{Fig2}b and c. The current $I_{\mbox{\tiny B}}$ ($I_{\mbox{\tiny D}}$) shows
predominantly the Coulomb maxima of the dot in segment~2 (segment~1) while those of the other dot are much weaker. The non-local voltage in Fig.~\ref{Fig2}(a) displays clear signatures of conductance resonances of both quantum dots. The lever arm of gate
voltage 1 (2) on the dot in segment 2 (1) is about a factor of six
smaller than for the direct gate voltage. There are stripes with
an intermediate slope in all three quantities which we attribute to the formation of standing waves between the two quantum dots near lead~A.
This resonator is coupled in series to both quantum dots, a fact which leads to a corresponding modulation of the dot currents in certain ranges of gate voltages and magnetic fields.

\begin{figure}[t]
\centering
\includegraphics[width=8.5cm]{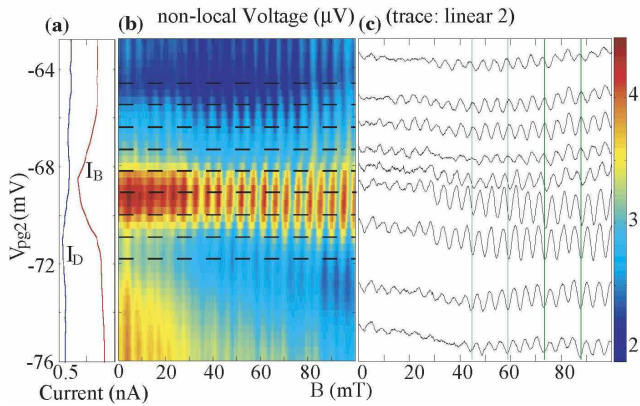}
\caption{\label{Fig3} AB oscillations in non-local voltage $V_\mathrm{nl}$ as function of the plunger gate voltages tuned along trace
`linear2'. (a) The currents $I_{\mbox{\tiny D}}$ and $I_{\mbox{\tiny B}}$ at a magnetic field of 95~mT. (b) Color plot of $V_\mathrm{nl}$ as function of plunger gates and magnetic field. (c) Cross-sections along dashed lines in (b) giving $V_\mathrm{nl}$ vs. magnetic field.
The dashed lines mark constant values of magnetic field where a $\pi$ phase shift of the AB
oscillations can be seen, as gate voltages are tuned
across the conductance peak in dot~2.}
\end{figure}

In order to study the phase evolution of the transmission through the quantum dots we measure $V_\mathrm{nl}$ as a function of magnetic field.
The results presented below were obtained in a
regime where the coupling to the leads was strong enough, but the dots were still well defined.
By tuning $V_{\mathrm{pg1}}$ and $V_{\mathrm{pg2}}$ we follow the trace labeled `linear2' in
Fig.~\ref{Fig2}(a), staying on a conductance maximum in dot~1
while stepping through a conductance peak in dot~2. As shown in Fig.~\ref{Fig3}(a) the current
$I_{\mbox{\tiny B}}$ shows the expected Coulomb peak of dot~2 as a function
of plunger gate voltage, while $I_{\mbox{\tiny D}}$
changes little.
At each gate voltage setting the AB oscillations are measured. The
non-local voltage in Fig.~\ref{Fig3}(b) shows a pronounced
AB signal which is strongest in the range of the conductance maximum.
A set of specific traces of $V_\mathrm{nl}$ versus magnetic field
is depicted in Fig.~\ref{Fig3}(c). The vertical lines connect minima in the lowest trace with maxima
in the uppermost trace. This corresponds to a phase shift of
$\pi$.

The multi-probe conductance formula~\cite{Buttiker86} gives for the
non-local voltage in our setup:
\[
V_{\mathrm{nl}}=-\frac{T_{\mbox{\tiny CA}}}{T_{\mbox{\tiny CC}}} V_{\mathrm{bias}}.
\]
The transmission coefficients $T_{ij}$ from probe~$j$ to $i$ are named
according to Fig.~\ref{Fig1}(b). Considering only oscillations
with a periodicity of $h/e$ we assume
$T_{\mbox{\tiny CC}}=-\sum_{i\neq\mbox{\tiny C}}T_{\mbox{\tiny C}i}$ to be constant. In this
case AB oscillations in $V_\mathrm{nl}$ reflect the interference contribution to $T_{\mbox{\tiny CA}}$ from which the phase difference between the two interfering transmission amplitudes can be directly read~\cite{Schuster97}.
In contrast, phase changes seen in the two currents
do not directly reflect the difference of transmission phases of the dots. In the following we
therefore consider only phase shifts of the AB oscillations in $V_\mathrm{nl}$.

\begin{figure}[t]
\centering
\includegraphics[width=8.5cm]{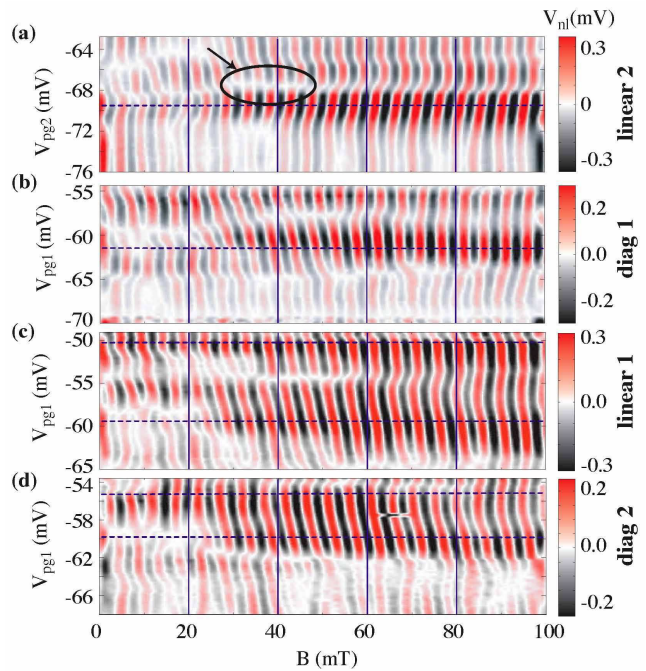}
\caption{\label{Fig4}
AB oscillations along the different traces indicated in Fig.~\ref{Fig2}. (a) sweep `linear 2' (b) sweep `diag 1' (c) sweep `linear 1' and (d) sweep `diag2'. Vertical lines are guides to the eye. Dashed horizontal lines indicate, where maxima in the magnetic field averaged contribution to $V_\mathrm{nl}$ occur.}
\end{figure}

We have analyzed such phase shifts quantitatively for all four
traces indicated by the arrows in Fig.~\ref{Fig2}(a). The fast Fourier transform of each magnetic field sweep was multiplied with the filter function $f(\omega)=(\sigma\omega)^2/2\exp[1-(\sigma\omega)^2/2]$ with $\sigma=h/(e\sqrt{2}\pi A)$ in order to obtain the pure $h/e$-periodic contribution. The inverse fast Fourier transform of the filtered data gives the oscillatory component of $V_\mathrm{nl}$ as a function of magnetic field which is plotted in Fig.~\ref{Fig4}. We have verified that this filtering procedure does not influence the phase of the AB oscillations by carefully comparing filtered data with raw data.

We begin the discussion with Fig.~\ref{Fig4}(a) corresponding to trace `linear2' in Fig.~\ref{Fig2}(a). The dashed horizontal line indicates the maximum in $V_\mathrm{nl}$ averaged over $B$. The AB amplitude is strongest close to this line, as expected, but displays a pronounced $B$-dependence, i.e. it is significantly larger for $B>30$\,mT than below. The phase shift  can be directly read from the shift of AB maxima or minima as a function of $V_\mathrm{pg2}$. The general trend is a shift to larger magnetic fields with increasing $V_\mathrm{pg2}$. The expected phase shift of $\pi$ across the conductance resonance is clearly discernible for $B>50$\,mT. However, there is a dip in the AB amplitude at $V_\mathrm{pg2}\approx -68$\,mV which has a strong phase shift associated with it. At $B\approx 40$\,mT and $V_\mathrm{pg2}\approx -68$\,mV [arrow in Fig.~\ref{Fig4}(a)] the AB amplitude is zero and a phase lapse of $\pi$ can be identified. This phase lapse lies in the flank of the peak of the field averaged $V_\mathrm{nl}$ and occurs at a point of zero AB amplitude, similar to the phase lapses observed in Ref.~\onlinecite{Schuster97} and in agreement with the general theoretical discussion of phase lapses in Ref.~\onlinecite{Buttiker99}. The novel aspect here is the occurrence of the phase lapse only in a limited range of magnetic field. Beyond this field range the phase lapse disappears in favor of a continuous phase evolution of order $\pi$.

Along trace `diag1' in Fig.~\ref{Fig2}(a) an
electron is added to both dots simultaneously.
In this case we expect a phase shift of $\pi-\pi=0$ since the phase difference between the arms of the ring remains constant.
In the experiment the phase shift 
 in Fig.~\ref{Fig4}(b) is indeed small for $B>50$\,mT, it amounts to about $\pi/10$. 
At lower fields, however, phase lapses and related strong phase shifts occur like that observed for trace `linear2'.

When we measure along trace `linear1'
[Fig.~\ref{Fig4}(c)] the gross trend of the phase shift differs in sign from that in Fig.~\ref{Fig4}(a), in agreement with the fact that we tune the dot in the other segment.
In the coupling regime of the quantum dots where AB oscillations are observable, Coulomb resonances can occur close to each other. This is the case for this sweep as indicated by the two dashed blue horizontal lines in Fig.~\ref{Fig4}(c).
In such a case the question arises how the transmission phases of individual resonances combine to the observed phase shift. In our measurement we find that the phase accumulation across the first resonance just above $V_\mathrm{pg1}=-60$\,mV is about $-\pi$ for $B$ between 30 and 70\,mT, as expected. At higher fields the phase shift weakens.
At $V_\mathrm{pg1}\approx -55$\,mV and between -40 and -50\,mT, phase lapses occur which influence the phase evolution at higher magnetic fields but cannot compensate the accumulated phase completely. At the top edge of the figure, the second resonance leads to a further phase accumulation in the same direction as the first. 

Along trace `diag2' [Fig.~\ref{Fig4}(d)] an electron is removed from dot~2
and added to dot~1. As expected, AB oscillation maxima shift in the same direction as in `linear1'. However, at $B=80$\,mT, for example, the phase accumulated between $V_\mathrm{pg1}=-68$\,mV and $-56$\,mV is only slightly more than $\pi$ rather than $2\pi$. Again, the phase evolution is influenced by occasional phase lapses.

Summarizing the observations we can make the following statements. We have succeeded in measuring the AB phase in a four terminal quantum ring with two dots embedded in two different segments. The AB oscillations tend to be well developed and suitable for phase measurements only at slightly elevated magnetic fields. In the range between 50 and 80\,mT the expected phase shifts across Coulomb-blockade resonances are typically observed. Phase lapses occur occasionally in finite magnetic field ranges at specific gate voltages. They are accompanied by a vanishing AB amplitude. Even outside the $B$-ranges where phase lapses occur, the phase evolution can be strongly modified by their presence. Following the argument of B\"uttiker and coworkers \cite{Buttiker99,Buttiker00}, the transmission amplitude follows a certain curve in the complex plane as a function of gate voltage. A phase lapse accompanied with a zero in transmission occurs when this line goes through the origin. A change in magnetic field may shift this curve a little leading to a strong, but continuous phase shift accompanied with small transmission. The appearance of phase lapses in finite magnetic field ranges impedes the definition of a magnetic field independent transmission phase. A conceivable origin of the phase lapses is the finite width of the ring segments accommodating several modes. In the magnetic field range investigated, the classical cyclotron radius is larger or comparable to the ring radius. A small influence of Lorentz force effects cannot be excluded at the highest fields.

Our samples differ from those in Ref.~\onlinecite{Schuster97}
by design and in view of the technological approach.
We have adopted the general idea of using reflecting walls to guide the electrons ballistically
around the ring without loosing too many of them into the
contacts. Compared to Ref.~\cite{Schuster97}
the most noticeable feature of our structure is the second dot in the reference arm. Its addition gives
individual control over the accumulated phases in both arms. Since the reference dot is kept on a conductance resonance, i.e. in the position of maximum slope of the phase evolution, while the other is swept through a resonance, the reference dot
is very sensitive to phase changes.
Should the reference dot be tuned slightly off its resonance during
the sweep `linear 2', one would expect a pronounced
influence on the measured phase evolution. The fact that we still
observe a phase change of about $\pi$ on trace `linear2', when dot~2 is tuned through a
conductance resonance, lets us conclude that the transmission phase
through dot~1 is rather stable and the observed phase shift is dominated by dot~2.

In this experiment, the transmission phase was measured in a system where two Coulomb blockaded quantum dots are embedded in a four-terminal quantum ring. We have shown
that the phase evolution can be determined with individual control over the
electron occupancy in each dot.
The experiments demonstrate that the measured transmission phase depends on the magnetic field range analyzed. Phase lapses can arise at certain magnetic fields, while at others they give way to a continuous phase evolution. In certain magnetic field ranges the phase behavior is in agreement with theoretical expectations, but its behavior is much more complex than anticipated.
This possibly suggests that structural resonances, in addition to the dots, play a role in the overall transmission phase of the system.
\begin{acknowledgments}
We thank M. B\"uttiker, Y. Gefen, M. Heiblum, H. Weidenm\"uller for valuable discussions.
Financial support by the Swiss Science Foundation (Schweizerischer
Nationalfonds) is gratefully acknowledged.
\end{acknowledgments}

\bibliography{apssamp}

\end{document}